\documentclass[onecolumn,11pt]{article}
\usepackage{graphics}
\usepackage[pdftex]{graphicx}
\usepackage{amssymb}
\usepackage{amsmath}
\usepackage{amsthm}
\usepackage{hyperref}
\newtheorem{theorem}{Theorem}

\newtheorem{lem}{Lemma}
\newtheorem{result}{Result}
\newtheorem{remark}{Remark}
\begin{document}
 \title{On the Population Size Estimation from Dual-record System: Profile-Likelihood Approaches}
 \author{Kiranmoy Chatterjee\thanks{Department of Statistics, Bidhannagar College, Kolkata-700064, India; E-mail: \emph{kiranmoy07@gmail.com}} \\
Diganta Mukherjee
\thanks{Sampling and Official Statistics Unit, Indian Statistical Institute, Kolkata-700108, India.}}

\date{}
 \maketitle

\begin{abstract}
Motivated by various applications, we consider the problem of homogeneous human population size ($N$) estimation from Dual-record system (DRS) (equivalently, two-sample capture-recapture experiment). The likelihood estimate from the independent capture-recapture model $M_t$ is widely used in this context though appropriateness of the behavioral dependence model $M_{tb}$ is unanimously acknowledged. Our primary aim is to investigate the use of several relevant pseudo-likelihood methods profiling $N$, explicitly for model $M_{tb}$. An adjustment over profile likelihood is proposed. Simulation studies are carried out to evaluate the performance of the proposed method compared with Bayes estimate suggested for general capture-recapture experiment by Lee et al. (Statistica Sinica, 2003, vol. 13). We also analyse the effect of possible model mis-specification, due to the use of model $M_t$, in terms of efficiency and robustness. Finally two real life examples with different characteristics are presented for illustration of the methodologies discussed.
\paragraph{}\emph{Key words:} Adjusted profile likelihood; Behavioral response; Model mis-specification; Modified profile likelihood; Nuisance parameters; Robustness.
\end{abstract}

\section{Introduction}\label{sec_intro}
 The problem of human population size estimation is a very important statistical concern which includes a vast area of application in the fields of epidemiology, demography and official statistics. Census or civil registration system often fails to extract the true size of the population. Usually they conduct another survey independently after the census operation to estimate the number of events missed in the census count. This is equivalent with capture-recapture principle for the estimation of true size, say \emph{N}, of the target population. Several likelihood models along with associated estimates from capture-recapture technique were first addressed by Otis et al. (1978 \cite{Otis78}) for different plausible situations with $T (\geq2)$, number of independent sources of information. Application of this technique for estimation of the number of affected people in an epidemiological study or in a particular event (like war, natural calamity, etc.) is also very popular in interdisciplinary platform. In the context of human population, more than two sources of information is hardly found for any problem.
\paragraph{}
Different models for population size estimation based on Dual-record system (DRS) have been well-sketched by Wolter (1986 \cite{Wolter86}). In practice for homogeneous group, model $M_t$ has received much attention from both the frequentist and Bayesian statisticians. $M_t$ accounts for time(\textit{t}) variation effect and assumes independence between the sources of information. This model was first analysed by Chandrasekar and Deming (1949 \cite{Chadra49}) for estimation of vital events for a human population. Various frequentist and likelihood approaches are present in the capture-recapture literature (see, Bishop et al. (1975 \cite{Bish75}), Huggins (1989 \cite{Hugg89})). Bayesian approach is pioneered by Robert (1967 \cite{Robert67}), Castledine (1981 \cite{Castle81}) and Smith (1988 \cite{Smith88}; 1991, \cite{Smith91}) and George and Robert (1990 \cite{George90}, Technical Report). George and Robert (1992 \cite{George92}) first gave an extensive account on the population size estimation through hierarchical Bayesian analysis via Gibbs sampling on model $M_t$. But this common model would not be appropriate in most of the situations for human population, especially when capture probabilities also vary with behavioral response. At the time of second capture, those who are caught in the first sample have a significant difference than those who are not captured previously. When both the time (\textit{t}) variation effect and behavior response (\textit{b}) effect acts together then we will have a more complicated model $M_{tb}$, where behavioral response effect is modelled by the parameter $\phi$. Particularly, when $\phi=1$, then $M_{tb}$ reduces to $M_{t}$. Otis et al. (1978 \cite{Otis78}) addressed the non-identifiability problem related to this model and Chao et al. (2000 \cite{Chao00}) derived \textit{mle} following Lloyd$'$s (1994 \cite{Lloyd94}) assumption only when $T\geq3$. Though the relevancy of the model $M_{tb}$ is understood in many situations, but due to lack of identifiability for DRS i.e. when $T=2$, $M_{tb}$ is seldom used for human population and model $M_t$ is widely employed for its simplicity in both demographic and epidemiological studies. Hence the issue of model mis-specification is raised. Lee and Chen (1998 \cite{Lee98}) and Lee et. al. (2003 \cite{Lee03}) successfully used the subjective Bayesian technique to $M_{tb}$ for $T\geq3$ through Gibbs sampling. Chatterjee and Mukherjee (2014 \cite{Chatterjee14}) discusses some issues related to the full Bayes method specifically for DRS and develops some empirical Bayes strategies considering the problem of $N$ estimation in a missing data framework. In Bayesian paradigm, difficulty may arise as the resulting estimator for $N$ may be very sensitive to the choice of prior(s).
\paragraph{}
Estimation of population size \emph{N} from $M_{tb}$ is the main interest of this article and another aim is to study the effect of model mis-specification due to the use of model $M_t$ even when $\phi$ is in a small neighbourhood of 1. Here, all the model parameters except $N$ are regarded as nuisance parameters. Some useful likelihood-based inference through the construction of pseudo-likelihood functions by eliminating the nuisance parameters are discussed in literature (\textit{see} Cox, 1975 \cite{Cox75}; Basu, 1977 \cite{Basu77}; Berger et al., 1999 \cite{Berger99}). As per our knowledge, profile and adjusted profile likelihood (Cox and Reid 1987 \cite{Cox-Reid87}) for model $M_t$ has been studied by Bolfarine et al. (1992, \cite{Bolfarine92}). Recently, Salasar et al. (2014 \cite{Salasar14}) analysed integrated likelihood approach, another pseudo-likelihood method, with uniform and Jeffrey's prior for eliminating nuisance parameters in $M_{t}$. However, in this article, we confine ourselves to the profile likelihood and some of its relevant modifications that can summarize the set of likelihoods $\{L(N,\psi|\underline{\textbf{x}}):\psi\in\Psi\}$ over $\Psi$. The goal of the article is to explicitly investigate the potential of these profile likelihood related methods for both the models $M_{tb}$ and $M_{t}$ in DRS context only. We also proposed an adjustment to the profile likelihood for the generic model $M_{tb}$. In summary, this article is framed to evaluate the extent of inefficiency in the simple estimate $\hat{N}_t$ and also to provide a non-Bayesian alternative for model $M_{tb}$.
\paragraph{}
In the next section, we discuss the models \emph{$M_{t}$} and \emph{$M_{tb}$} in DRS context. Performance of the widely used estimate $\hat{N}_{ind}$ from model $M_t$ is analysed in terms of bias and variance when independence assumption is violated due to behavioral response change. In section 3, the profile and modified profile likelihood functions are discussed with implementations to our interest models. Therefrom, we develop an adjustment to the profile likelihood for $M_{tb}$ in section 4. Evaluation of the proposed adjusted profile likelihood approach is carried out by an extensive simulation study in section 5 and comparison made with Bayes estimate sketched by Lee et al. (2003 \cite{Lee03}). Comparative graphical investigations on the performance and robustness of the proposed approach are done against the common estimate $\hat{N}_{ind}$. Then, illustration of our method is discussed through the application to real datasets and finally in section 6, we summarize our findings and provide some comments about the usefulness of above profile likelihood based approaches.

\section{Dual Record System: Preliminaries}\label{sec_drs}
Let us consider a given human population \emph{U} whose size \emph{N} is to be estimated and any attempt to enlist all individuals in $U$ is believed to be incomplete as it fails to capture all. To have better estimate of true \emph{N}, minimum two sources of information covering that population is needed. In this paper we will concentrate on those models which have two common assumptions - (1) population is closed within the time of two different sources, (2) individuals are homogeneous with respect to capture probabilities in both the sources. When information is collected from two sources, it is known as Dual-record System (DRS). The individuals captured in first source (list 1) are matched with the list of individuals from second source (list 2). Classify all the captured individuals in \emph{U} according to a multinomial fashion as in Table 1. The total number of distinct captured individuals by the two lists is $x_0$ (say), then $x_0=x_{10}+x_{01}+x_{11}$. Clearly, the number of missed individuals $x_{00}$ by both systems is unknown and that makes the total population size \emph{N}($=x_{..}$) unknown. Expected proportion or probability associated with each cell are also given and these notations will be followed throughout in this paper.
\begin{table}[h]
\begin{center}
\caption{$2\times2$ table for Dual-System Model}
\begin{tabular}{lccc}
\hline
&\multicolumn{3}{c}{List 1} \\
\cline{2-4}
List 2 & In & out & Total\\
\hline \hline
&\multicolumn{3}{c}{I. Observed sample numbers} \\
In & $x_{11}$ & $x_{01}$ & $x_{.1}$\\
Out& $x_{10}$ & $x_{00}$ & $x_{.0}$\\ 
\hline
Total& $x_{1.}$ & $x_{0.}$ & $x_{..}=N$\\
\hline
&\multicolumn{3}{c}{II. Expected Proportions} \\
In & $p_{11}$ & $p_{1.}$ & $p_{.1}$\\
Out& $p_{.1}$ & $p_{00}$ & $p_{.0}$\\ 
\hline
Total& $p_{1.}$ & $p_{0.}$ & 1\\
\hline
\end{tabular}
\end{center}
\label{tab:dse}
\end{table}
Combining all the information estimate of \emph{N} could be obtained assuming different conditions on the individual's capture probabilities leading to different models. In this article, we confine ourselves to the models $M_t$ and $M_{tb}$ which are appropriate for homogeneous human population or sub-population.

\subsection{Model $M_{t}$}\label{subsec_model_Mt}
This model is very simple and widely used for human population. Two additional assumptions are required for this model. One is that the two lists are \textit{causally independent}. An individual being included in List 2 is independent of his/her inclusion in List 1. Another is \textit{time variation} in the capture probabilities, i.e., two marginal capture probabilities satisfy $p_{1.}\neq p_{.1}$. Then the associated likelihood for $N(\geq x_0)$ is\\
\[ L_{t}(N,p_{1.},p_{.1})=
      \frac{N!}{x_{11}!x_{01}!x_{10}!(N-x_0)!}p_{1.}^{x_{1.}}p_{.1}^{x_{.1}}(1-p_{1.})^{N-x_{1.}}(1-p_{.1})^{N-x_{.1}}.\]
The corresponding maximum likelihood estimates are \\
\[ \hat{N}_{t} = x_{11}+x_{01}+x_{10}+\left[\frac{x_{01}x_{10}}{x_{11}}\right]=\left[\frac{x_{.1}.x_{1.}}{x_{11}}\right], \]
\[ \hat{p}_{01,t} =\frac{x_{11}}{x_{.1}} \hspace{0.2in} \mbox{and} \hspace{0.2in} \hat{p}_{10,t}=\frac{x_{11}}{x_{1.}}. \]
This estimator is well-known as DSE or C-D estimator in the literature of census coverage error estimation and it also popular as Lincoln-Petersen estimator in wildlife population study. We denote this estimator as $\hat{N}_{ind}$ throughout this paper.

\subsection{Model $M_{tb}$}\label{subsec_model_Mtb}
Causal independence assumption is criticised in surveys and censuses of human populations. An individual who is captured in first attempt may have more (or less) chance to include in the second list than the individual who has not been captured in first attempt. This change in behavior may occur due to different causes (\textit{see} Wolter 1986 \cite{Wolter86}) and it is grossly known as behavioral response variation. When this chance is more then the corresponding individuals are treated as \emph{recapture prone}, otherwise when it is less, the individuals become \emph{recapture averse}. When this feature is combined with the \textit{time variation} assumption, one will get the relatively complex model $M_{tb}$. To model this situation one has to impose the assumption following Wolter (1986 \cite{Wolter86}) that the probability of first capture is the same for each individual in the population and that is
\[\mbox{Prob(\emph{i}th individual is captured in List 1) = $p_{1.}$}.\]
\[\mbox{Prob(\emph{i}th individual is captured in List 2 $|$ not captured in List 1) = $p_{01}/p_{0.}$ = $p$}\]
and the probability of recapture or Prob(\emph{i}th individual is captured in List 2 $|$ he/she is captured in List 1)= $p_{11}/p_{1.}$ = $c$.
But this model has some unidentifiability issue as the corresponding likelihood function
\begin{center}
\begin{eqnarray}
L_{tb}(N,p_{1.},p,c) &\propto& \frac{N!}{(N-x_0)!}c^{x_{11}}p_{1.}^{x_{1.}}p^{x_{01}}(1-p_{1.})^{N-x_{1.}}(1-p)^{N-x_{0}}(1-c)^{x_{10}},\label{Eq_L_1}
\end{eqnarray}
\end{center}
for $N>x_0$, consists lesser number of sufficient statistics ($x_{11},x_{01},x_{10}$) than the parameters ($N,p_{1.},p,c$) (\textit{see} Otis et al. 1978 \cite{Otis78}). A popular assumption that recapture probability at second sample, \emph{c}, is equal to a constant multiple of the probability of first time capture in second attempt, $p$. Hence, $c=\phi p$ and Chao et al. (2000 \cite{Chao00}) adopted this from Lloyd (1994 \cite{Lloyd94}) to get rid of from the problem . Then likelihood becomes
\begin{center}
\begin{eqnarray}
L_{tb}(N,p_{1.},p,\phi) &\propto& \frac{N!}{(N-x_0)!}\phi^{x_{11}}p_{1.}^{x_{1.}}p^{x_{.1}}(1-p_{1.})^{N-x_{1.}}(1-p)^{N-x_{0}}(1-\phi p)^{x_{10}}\label{Eq_L_2}
\end{eqnarray}
\end{center}
 where $\phi$, the behavioral response effect, is orthogonal to $N$. Lloyd's assumption is helpful when number of sources is strictly more than two.
But it is noticed that identifiability problem persists in DRS. Both of $\phi$ and $p$ are not identifiable separately but their product $c$ is rather identifiable. Thus, likelihood (\ref{Eq_L_2}) is more ill-behaved than (\ref{Eq_L_1}). Replacing $p$ with $c/\phi$ in (1) one might have another parametrization where $\phi$ is not at all orthogonal to $N$.

\subsection{Model Mis-specification}\label{subsec_model_mis}
In the context of several real life applications on homogeneous human population or subpopulations, estimator $\hat{N}_{ind}$ derived from model $M_t$ is often used though appropriateness of model $M_{tb}$ is well-understood. Hence, a threat of model mis-specification naturally arises if $\hat{N}_{ind}$ is used. In this section we investigate how serious that threat could be. At first, consider the following lemma (\textit{see} Raj, 1977 \cite{Raj77}).
\begin{lem}
Suppose $x, y$ and $z$ are three random variables with finite moments
upto second order. Then, large sample approximation to the mean of $\frac{xy}{z}$ is
\[\emph{E}\left(\frac{xy}{z}\right) \approx \frac{\emph{E}(x)\emph{E}(y)}{\emph{E}(z)}\left(
1+\frac{C(x,y)}{E(x)E(y)}-\frac{C(x,z)}{E(x)E(z)}-\frac{C(y,z)}{E(y)E(z)}+\frac{V(z)}{E^2(z)}\right)\]
\end{lem}
Replacing $x$, $y$ and $z$ by $x_{1.}$, $x_{.1}$ and $x_{11}$ respectively in the above lemma, we obtain the bias stated in the next theorem. Variance is also computed using same lemma with suitable replacement.
\begin{theorem}\label{Theo_1}
Suppose, the actual underlying model is $M_{tb}$ with parametrization ($N,p_{1.},p,\phi$). Then, second order large sample approximation to the bias and variances of $\hat{N}_{ind}=\left(\frac{x_{1.}x_{.1}}{x_{11}}\right)$ for estimating $N$ are
\begin{center}
\begin{eqnarray}
Bias\left(\hat{N}_{ind}\right)_{M_{tb}}&=&N(1-p_{1.})\frac{1-\phi}{\phi}+\frac{1}{\phi}\frac{(1-p_{1.})(1-\phi p)}{p_{1.}\phi p},\nonumber\\
Var\left(\hat{N}_{ind}\right)_{M_{tb}}&=&N\frac{1}{\phi}\frac{(1-p_{1.})(1-\phi p)}{p_{1.}\phi p}.\nonumber
\end{eqnarray}
\end{center}
\end{theorem}
Clearly, when $\phi$ increases above one, second part of the right hand side in bias gradually boils down to 0 as $p_{1.}$ and $\phi p=c$ are expected to be more than 0.5. Hence, simple estimate $\hat{N}_{ind}$ underestimates $N$ and its bias $\rightarrow -N(1-p_{1.})$ as $\phi$ ($>1$) increases. Similarly, when $\phi$ ($<1$) decreases to $0$, $\hat{N}_{ind}$ increasingly overestimates $N$. Thus, assumption of $\phi=1$ might happen to be very risky and use of $\hat{N}_{ind}$ may lead to an inefficient estimate. On the other hand, if $\phi$ is exactly 1 (i.e. list-independence case), bias reduces to $\frac{(1-p_{1.})(1-p_{.1})}{p_{1.}p_{.1}}$, as $p=p_{.1}$ under independence. Therefore, bias will be negligible when $p_{1.}$ and $p_{.1}$ both are large. The result also tells that s.e.($\hat{N}_{ind}$) is proportional to $O(N^{1/2})$ under $M_{tb}$. Even when, $\phi=1$, then
\[s.e.\left(\hat{N}_{t}\right)_{M_t}=N^{1/2}\left\{\frac{(1-p_{1.})(1-p_{.1})}{p_{1.}p_{.1}}\right\}^{1/2}=O(N^{1/2}).\]
\paragraph{}
Our discussion on pseudo-likelihood methods in next two sections is based on both the models $M_{tb}$ and $M_{t}$, since, model $M_{t}$ is often used in practice and $M_{tb}\equiv M_{t}$ only when $\phi=1$.

\section{Some Pseudo-likelihood Methods}\label{sec_pseudo}
Let us consider a statistical model with likelihood function $L(\lambda|\underline{\textbf{x}})$ with $\lambda=(\theta, \psi)$, where $\theta$ is parameter of interest and $\psi$ represents nuisance parameter, both may be vector valued. Presence of
more nuisance parameters in the model affects the comparative inferential study based on the likelihood (\textit{see} Basu (1977 \cite{Basu77}), Severini (2000 \cite{Severini00})). Now our aim is to find a function that can summarize the set of likelihoods $\mathcal{L}^{*}=\{L(\theta,\psi|\underline{\textbf{x}}):\psi\in\Psi\}$ over $\Psi$. That summarized function is denoted as $L^*(\theta)$ which is treated some what like a likelihood function of $\theta$; as if the inference frame has $\theta$ as the only parameter. We refer such functions $L^*(\theta)$ here as pseudo likelihood function of $\theta$. This kind of pseudo likelihood functions includes profile likelihood function. Modified profile likelihoods (Barndorff-Nielsen, 1983 \cite{Barndorff-nielsen83} and 1985 \cite{Barndorff-nielsen85}) and adjusted profile likelihoods (Cox and Reid, 1987 \cite{Cox-Reid87}) are basically modifications to the profile likelihood function. There are several other kind of pseudo likelihood functions in the literature, such as marginal, conditional, partial (Cox, 1975 \cite{Cox75}) and integrated likelihood (Berger et al., 1999 \cite{Berger99}) functions. In the present context, interest is basically on $N$ and sometimes also on $\phi$ in $M_{tb}$. We restrict ourselves to the profile likelihood functions obtained by summarising the original data likelihood over the domain of nuisance parameter and some of its suitable modifications. Moreover, we propose an adjustment over profile likelihood which is driven by an adjustment coefficient so that the resulting likelihood estimate satisfies some desirable frequentist properties.

\subsection{Profile Likelihood (PL) Method}\label{subsec_method_PL}
This approach summarizes $\mathcal{L}^{*}$ at $\psi=\hat{\psi}_{\theta}$, the conditional \textit{mle} of $\psi$ for given $\theta$. Hence, the profile likelihood (PL) for $\theta$ is $L^{P}(\theta)=L(\theta,\hat{\psi}_{\theta}|\underline{\textbf{x}})$. Hence, inference about $\theta$ is made by maximizing $L^{P}(\theta)$ (or $logL^{P}(\theta)$) considering as a likelihood function (or log-likelihood function) of $\theta$. But, in general, it is not a proper likelihood function. Thus, inferences based on this assumption may be misleading, specifically when $\psi$ is high-dimensional.
\paragraph{}
In the context of independent model, $M_{t}$, PL for interest parameter $N$ is given by
\[L_{t}^{P}(N)=\frac{N!}{(N-x_0)!}(N-x_{1.})^{(N-x_{1.})}(N-x_{.1})^{(N-x_{.1})}N^{-2N},\]
for $N\geq max(x_{1.},x_{1.},x_{0})=x_0$. Here, as elsewhere in the paper, multiplicative terms not depending on $N$ in likelihood function of $N$ have
been ignored.
\begin{theorem}\label{Theo_2}
$L_{t}^{P}(N)$ is increasing in $N$ for $N<(x_{1.}x_{.1}/x_{11})-1$ and hence, when $(x_{1.}x_{.1}/x_{11})$ is an integer, the corresponding \textit{mle} $\hat{N}_{t}^{P}$ is $(x_{1.}x_{.1}/x_{11})-1$. When $(x_{1.}x_{.1}/x_{11})$ is not an integer, $\hat{N}_{t}^{P}$ is either $[x_{1.}x_{.1}/x_{11}]-1$ or $[x_{1.}x_{.1}/x_{11}]$, according to which produces the maximum value of the profile likelihood,
where $[u]$ denotes the greatest integer not greater than $u$, for $u\in\mathbb{R}$.
\end{theorem}
$\hat{N}_{t}^{P}$ is finite iff $x_{11}>0$. Maximum profile likelihood (PL) estimate can also be obtained by maximising $L_{t}^{P}(N)$ assuming $N$ as a real number and using the formula for \textit{digamma function} of any positive integer $z$ (obtained from recursion relation), $\beta(z)=(\partial/\partial z) log(\Gamma(z))=-\gamma+\Sigma_{a=1}^{z-1}(1/a)$, where $\gamma$ is the \textit{Euler-Mascheroni constant}.
\paragraph{}
For any parametrization of model $M_{tb}$, such as (\ref{Eq_L_1}) or (\ref{Eq_L_2}), the PL for $N$ reduces to
\[L_{tb}^{P}(N)=\frac{N!}{(N-x_0)!}(N-x_{0})^{(N-x_{0})}N^{-N},\]
for $N>x_{0}$, as PL is parametrization invariant. Clearly $L_{tb}^{P}(N)$ is decreasing for $N>x_{0}$ as $\prod_{i=1}^{x_0-1}(1-\frac{i}{N})<(1-\frac{1}{N})^{x_0-1}$. It can be written that $L_{tb}^{P}(N)=(1-\frac{x_0}{N})^{N-x_0}\prod_{i=1}^{x_0-1}(1-\frac{i}{N})<(1-\frac{1}{N})^{N-1}$. Now as $(1-\frac{1}{N})^{N-1}\downarrow N$, $L_{tb}^{P}(N)$ is a decreasing function in $N$ for $N>x_{0}$. Hence, \textit{mle} will be the lower bound of $N$ i.e. $\hat{N}_{tb}^{P}=(x_0+1)$.
It is clear that this pseudo-likelihood is not useful, as it stands, for estimating the population size $N$.

\subsection{Modified Profile Likelihood (MPL) and Its Approximation (AMPL)}\label{subsec_method_MPL}
Since marginal and conditional likelihoods are not available for $M_{tb}$, the idea is to use a suitable modification to the profile likelihood. Several such modifications are suggested in the literature. PL cannot approximate a marginal or conditional likelihood function and that leads to poor performance. We now discuss a modification to the profile likelihood function. In general, modified profile likelihood (MPL) proposed by Barndorff-Nielsen (1983 \cite{Barndorff-nielsen83}, 1985 \cite{Barndorff-nielsen85}) is written as
\begin{center}
\begin{eqnarray}
L^{MP}(\theta)=D(\theta)|\hat{j}_{\psi\psi}(\theta,\hat{\psi}_{\theta})|^{-1/2}L^{P}(\theta)\label{Eq_MPL_1}.
\end{eqnarray}
\end{center}
where $D(\theta)=|\frac{\partial\hat{\psi}_{\theta}}{\partial\hat{\psi}}|^{-1}$, the inverse of jacobian $J(\theta)=\partial\underline{\textbf{x}}/\partial\hat{\psi}_{\theta}\propto\partial\hat{\psi}/\partial\hat{\psi}_{\theta}$ and $\hat{j}_{\psi\psi}$ is the observed Fisher information of $\psi$ for fixed $\theta$. The actual derivation of $L^{MP}(\theta)$ as an approximation to a conditional likelihood is sketched in Severini (2000 \cite{Severini00}) considering ($\hat{\psi}_{\theta},a$) as sufficient with $\theta$ held fixed and \textit{a} is ancillary statistic. However, we can simply express the partial derivative factor in $L^{MP}(\theta)$ as follows:
\paragraph{}
Let us denote the logarithm of likelihood $L(\cdot)$ as $\ell(\cdot)$. Then conditional \textit{mle} $\hat{\psi}_{\theta}$ implies $\frac{\partial\ell(\theta,\psi|\hat{\theta},\hat{\psi},a)}{\partial\psi}|_{\psi=\hat{\psi}_{\theta}}=0$, as sufficient statistics may be written as ($\hat{\theta},\hat{\psi},a$), $a$ being ancillary. Then, by differentiating with respect to $\hat{\psi}$ we have
\[\ell_{\psi;\psi}(\theta,\hat{\psi}_{\theta})\frac{\partial\hat{\psi}_{\theta}}{\partial\hat{\psi}}+\ell_{\psi;\hat{\psi}}(\theta,\hat{\psi}_{\theta})=0.\]
This implies $\frac{\partial\hat{\psi}_{\theta}}{\partial\hat{\psi}}$ = $\hat{j}_{\psi\psi}(\theta,\hat{\psi}_{\theta})^{-1}\ell_{\psi;\hat{\psi}}(\theta,\hat{\psi}_{\theta}),$ where $\hat{j}_{\psi\psi}(\theta,\hat{\psi}_{\theta})=-\ell_{\psi;\psi}(\theta,\hat{\psi}_{\theta})$. Hence, MPL in (\ref{Eq_MPL_1}) may also be written in the following form
\begin{center}
\begin{eqnarray}
L^{MP}(\theta)=|\ell_{\psi;\hat{\psi}}(\theta,\hat{\psi}_{\theta})|^{-1}|\hat{j}_{\psi\psi}(\theta,\hat{\psi}_{\theta})|^{1/2}L^{P}(\theta)\label{Eq_MPL_2},
\end{eqnarray}
\end{center}
and hence in (\ref{Eq_MPL_2}), $D(\theta)=|\hat{j}_{\psi\psi}(\theta,\hat{\psi}_{\theta})|/|\ell_{\psi;\hat{\psi}}(\theta,\hat{\psi}_{\theta})|$ according to the form in (\ref{Eq_MPL_1}).
\paragraph{}
There is an approximation to $L^{MP}$ suggested by Severini (1998 \cite{Severini98}) in which $D(\theta)$ is taken as $|\hat{j}_{\psi\psi}(\theta,\hat{\psi}_{\theta})|/|I(\theta,\hat{\psi}_{\theta};\hat{\theta},\hat{\psi})|$, where Fisher's information $I(\theta,\psi;\theta_0,\psi_0)=(\partial/\partial\psi_0)E\{\ell_{\psi}(\theta,\psi)|\theta_0,\psi_0\}$ is an approximation to $\ell_{\psi;\psi_0}(\theta,\psi)$ as $E\{\ell_{\psi}(\theta,\psi|\theta_0,\psi_0\}=\ell_{\psi}(\theta,\psi|\theta_0,\psi_0)+O(1)$ and $\ell_{\psi;\psi_0}(\theta,\psi)=(\partial/\partial\psi_0)\ell_{\psi}(\theta,\psi|\theta_0,\psi_0)$. Hence, approximated modified profile likelihood (AMPL) is
\begin{center}
\begin{eqnarray}
\widetilde{L}^{MP}(\theta)=|I(\theta,\hat{\psi}_{\theta};\hat{\theta},\hat{\psi})|^{-1}|\hat{j}_{\psi\psi}(\theta,\hat{\psi}_{\theta})|^{1/2}L^{P}(\theta)\label{Eq_MPL_3}.
\end{eqnarray}
\end{center}
\paragraph{}
\begin{remark}\label{rem_1}
Clearly, $L^{MP}(\theta)=\widetilde{L}^{MP}(\theta)$ if and only if $|\ell_{\psi;\hat{\psi}}(\theta,\hat{\psi}_{\theta})|=|I(\theta,\hat{\psi}_{\theta};\hat{\theta},\hat{\psi})|$, ignoring the terms not depending on $\theta$.
\end{remark}

\subsubsection*{\textit{Implementation to models $M_t$ and $M_{tb}$:}}\label{subsubsec_imple_method_MPL}
The following result shows that MPL and AMPL are identical on the domain $N\geq x_0$ for model $M_t$. Severini (1998 \cite{Severini98}) stated this result only. However, the explicit proof is given in \textit{Appendix}.
\begin{result}\label{Res_1}
Both $L^{MP}$ and $\widetilde{L}^{MP}$ are same for model $M_t$ with $\theta=N$, $\psi=(p_{1.},p_{.1})$ and for $N\geq x_0$, it is given by
\begin{center}
\begin{eqnarray}
L_{t}^{MP}(N)=\widetilde{L}_{t}^{MP}(N|\underline{\textbf{x}})&=&
\frac{N!}{(N-x_0)!}(N-x_{1.})^{(N-x_{1.}+1/2)}(N-x_{.1})^{(N-x_{.1}+1/2)}N^{-(2N+1)}\nonumber\\
&=&L_{t}^{P}(N)(N-x_{1.})^{1/2}(N-x_{.1})^{1/2}N^{-1}.\nonumber
\end{eqnarray}
\end{center}
\end{result}
\paragraph{}
An interesting relation between PL and MPL for the model $M_{t}$ is formulated in the next theorem. Theorem \ref{Theo_4} shows that MPL estimate is same as ordinary likelihood estimate of $N$. Proofs of the following two theorems are also in \textit{Appendix}.
\begin{theorem}\label{Theo_3}
The maximum profile likelihood estimator, $\hat{N}_{t}^{P}$, is no greater than the maximum modified
profile likelihood estimator $\hat{N}_{t}^{MP}$.
\end{theorem}
\begin{theorem}\label{Theo_4}
$L_{t}^{MP}(N)$ is increasing in $N$ for $N<(x_{1.}x_{.1}/x_{11})-1$ and hence, the corresponding \textit{mle}, $\hat{N}_{t}^{MP}$ is $[x_{1.}x_{.1}/x_{11}]$ if $(x_{1.}x_{.1}/x_{11})$ is not an integer; and is $(x_{1.}x_{.1}/x_{11})-1$, if $(x_{1.}x_{.1}/x_{11})$ is an integer.
\end{theorem}
\paragraph{}
Thus, for $(x_{1.}x_{.1}/x_{11})\in \mathbb{Z}^{+}$, the set of positive integers, $\hat{N}_{t}^{P}=\hat{N}_{t}^{MP}=\widetilde{N}_{t}^{MP}=(x_{1.}x_{.1}/x_{11})-1$ and for ($x_{1.}x_{.1}/x_{11}$) not $\in \mathbb{Z}^{+}$, $\hat{N}_{t}^{MP}=\widetilde{N}_{t}^{MP}=[x_{1.}x_{.1}/x_{11}]\geq\hat{N}_{t}^{P}$.
\paragraph{}
Next, we present the computation of MPL and AMPL in the context of model $M_{tb}$. Let us consider the parametrization $\theta=N$, $\psi=(p_{1.},p^{*}_{10},c)$. Hence, by differentiating the log-likelihood (from (\ref{Eq_L_1})) with respect to $\psi$, we have
$\ell_{\psi}^{tb}(\theta,\psi)=\left(\frac{x_{1.}}{p_{1.}}-\frac{N-x_{1.}}{1-p_{1.}},\frac{x_{01}}{p^{*}_{10}}-\frac{N-x_{0}}{1-p^{*}_{10}},\frac{x_{11}}{c}-\frac{x_{10}}{1-c}\right)$. Therefore,$E\{\ell_{\psi}^{tb}(\theta,\psi);\theta_0,\psi_0\}\left|_{\theta_0=\hat{\theta},\psi_0=\hat{\psi}}\right.=$\\
\\$\left(\frac{\hat{N}\hat{p}_{01}}{p_{1.}}-\frac{N-\hat{N}\hat{p}_{01}}{1-p_{1.}},\frac{\hat{N}\hat{p}^{*}_{10}(1-\hat{p}_{01})}{p^{*}_{10}}-\frac{N-\hat{N}\hat{p}^{*}_{10}(1-\hat{p}_{01})-\hat{N}\hat{p}_{01}}{1-p^{*}_{10}},\frac{\hat{N}\hat{c}\hat{p}_{01}}{c}-\frac{\hat{N}(1-\hat{c})\hat{p}_{01}}{1-c}\right)$.\\
\\Hence, $|I^{tb}(\theta,\hat{\psi}_{\theta};\hat{\theta},\hat{\psi})|\propto N^2(N-x_{1.})/(N-x_{0})$, since
\begin{center}
\begin{eqnarray}
I^{tb}(\theta,\psi;\hat{\theta},\hat{\psi})&=&\frac{\partial}{\partial\hat{\psi}} E\{\ell_{\psi}^{tb}(\theta,\psi);\theta_0,\psi_0\}\left|_{\theta_0=\hat{\theta},\psi_0=\hat{\psi}}\right.\nonumber\\
&=&\left(\begin{tabular}{ccc}
         $\frac{\hat{N}}{p_{1.}}+\frac{\hat{N}}{1-p_{1.}}$ & $-\frac{\hat{N}\hat{p}^{*}_{10}}{p^{*}_{10}}-\frac{\hat{N}\hat{p}^{*}_{10}-\hat{N}}{1-p^{*}_{10}}$&
         $\frac{\hat{N}\hat{c}}{c}-\frac{\hat{N}(1-\hat{c})}{1-c}$ \nonumber\\
         0 & $\frac{\hat{N}(1-\hat{p}_{01})}{p^{*}_{10}}-\frac{-\hat{N}(1-\hat{p}_{01})-\hat{N}\hat{p}_{01}}{1-p^{*}_{10}}$ & 0\nonumber\\
         0 & 0 & $\frac{\hat{N}\hat{p}_{01}}{c}+\frac{\hat{N}\hat{p}_{01}}{1-c}$\nonumber\\
         \end{tabular}
   \right)\\
\mbox{and $\hat{\psi}_{\theta}$}&=&\left(\frac{x_{1.}}{N},\frac{x_{01}}{N-x_{1.}},\frac{x_{11}}{x_{1.}}\right)\nonumber
\end{eqnarray}
\end{center}
Again, we have $|\ell_{\psi;\hat{\psi}}(\theta,\hat{\psi}_{\theta})|$ = $|\hat{j}_{\psi\psi}(\theta,\hat{\psi}_{\theta})|\left|\frac{\partial\hat{\psi}_{\theta}}{\partial\hat{\psi}}\right|$ and
$\left|\frac{\partial\hat{\psi}_{\theta}}{\partial\hat{\psi}}\right|=N^{-1}(N-x_{1.})^{-1}$ and $|\hat{j}_{\psi\psi}(\theta,\hat{\psi}_{\theta})|$ is found as $N^{3}(N-x_{1.})^{2}(N-x_0)^{-1}$.
Hence, $|\ell_{\psi;\hat{\psi}}^{tb}(\theta,\hat{\psi}_{\theta})|\propto N^2(N-x_{1.})/(N-x_{0})$, where
\begin{center}
\begin{eqnarray}
\ell_{\psi;\hat{\psi}}^{tb}(\theta,\hat{\psi}_{\theta})&=&\frac{\partial}{\partial\psi_0} \ell_{\psi}^{tb}(\theta,\psi|\theta_0,\psi_0)\left|_{\theta_0=\hat{\theta},\psi_0=\hat{\psi},\psi=\hat{\psi}_\theta}\right.\nonumber
\end{eqnarray}
\end{center}
So, we have $|\ell_{\psi;\hat{\psi}}^{tb}(\theta,\hat{\psi}_{\theta})|=|I^{tb}(\theta,\hat{\psi}_{\theta};\hat{\theta},\hat{\psi})|$, ignoring the terms not depending on $\theta=N$. Therefore, from (\ref{Eq_MPL_2}) and (\ref{Eq_MPL_3}), $L_{tb}^{MP}(N)=\widetilde{L}_{tb}^{MP}(N)$ and hence, the following result.
\begin{result}\label{Res_2}
For the model $M_{tb}$ with $\theta=N$, $\psi=(p_{1.},p^{*}_{10},c)$, both of $L_{tb}^{MP}$ and $\widetilde{L}_{tb}^{MP}$ is equivalent to
\begin{center}
\begin{eqnarray}
L_{tb}(N)&=&\frac{N!}{(N-x_0)!}(N-x_{0})^{(N-x_{0}+1/2)}N^{-(N+1/2)}\nonumber\\
&=&\mbox{$L_{tb}^{P}(N)(1-x_0/N)^{1/2}$, for $N>x_0$.}\nonumber
\end{eqnarray}
\end{center}
\end{result}
Now, $(\partial/\partial N)\ell_{tb}^{MP}(N)=(\partial/\partial N)\ell_{tb}^{P}(N)+\frac{1}{2(N-x_0)}-\frac{1}{2N}$.
Using the asymptotic approximation of gamma function, $log(\Gamma(z+1))=z\{log(z)-1\}+log(z)/2+log(2\pi)/2+O(z^{-1})$, we have $(\partial/\partial N)\ell_{tb}^{P}(N)=\frac{1}{2N}-\frac{1}{2(N-x_0)}+O(N^{-3})=O(-N^{-2})<0$ for $N>x_0$.
Therefore, $(\partial/\partial N)\ell_{tb}^{MP}(N)=(\partial/\partial N)\ell_{tb}^{P}(N)+\frac{x_0}{2N(N-x_0)}=O(N^{-3})>0$ for $N>x_0$. Hence clearly, $L_{tb}^{MP}$ also does not give any finite maximum likelihood estimate.
\paragraph{}
So far we have understood that $M_{tb}$ is the most suitable underlying model that a homogeneous capture-recapture system must follow and also the failure of this model even in case of modified and approximate modified profile likelihoods. That may lead the practitioners to use the model $M_t$ (assuming list-independence) whose \textit{mle} and other profile likelihoods exist. Here, in this paper, we try to address how much efficiency we are loosing by the use of $\hat{N}_{ind}$ if list-independence does not hold. The possible threat of model mis-specification due to the use of $M_t$ is discussed in section \ref{subsec_model_mis}. In the next section, we propose a suitable adjustment to the profile likelihood function for model $M_{tb}$ and discuss the conditions under which the associated estimate of $N$ can exist. The adjustment is so designed as to preserve better frequentist and robust properties than $\hat{N}_{ind}$ even in a small neighbourhood around 1.

\section{Inference Based on An Adjustment to Profile Likelihood (AdPL)}\label{sec_Infer_AdPL}
\subsection{Proposed Methodology for $M_{tb}$ and Related Properties}\label{subsec_method_AdPL}
Understanding the failure of PL and its two modifications - MPL and AMPL, for $M_{tb}$ here we propose an adjusted version of the profile likelihood. Our proposed adjusted profile likelihood (AdPL) for generic model $M_{tb}$ with adjustment coefficient $\delta$ ($\in\mathcal{R}$) is
\begin{eqnarray}
\widehat{L}^{AP}(\theta)=\left|\frac{\partial\hat{\psi}_{\theta}}{\partial\hat{\psi}}\right|^{-\delta}|\hat{j}_{\psi\psi}(\theta,\hat{\psi}_{\theta})|^{-1/2}L^{P}(\theta),\label{Eq_MPL_4}
\end{eqnarray}
Note that in particular, when $\phi=1$, $M_{tb}\Rightarrow M_{t}$ and therefore, $\widehat{L}^{AP}(\theta)$ will be same as $L^{MP}(\theta)$ in (\ref{Eq_MPL_1}) iff the adjustment coefficient $\delta$ is fixed at 1. That means, for model $M_{t}$, our proposed AdPL reduces to the MPL, $L_{t}^{MP}(N)$, given in \textit{Result} \ref{Res_1}, if $\delta=1$.
\paragraph{}
In the context of model $M_{tb}$ with parametrization (\ref{Eq_L_1}), $\left|\frac{\partial\hat{\psi}_{\theta}}{\partial\hat{\psi}}\right|=N^{-1}(N-x_{1.})^{-1}$.
Hence we have the following result using (\ref{Eq_MPL_4}) and $L_{tb}^{MP}(N)$.
\begin{result}\label{Res_3}
For model $M_{tb}$ with $\theta=N$, $\psi=(p_{1.},p^{*}_{10},c)$, the adjusted profile likelihood for all $N>x_0$, according to (\ref{Eq_MPL_4}), is given by
\begin{center}
\begin{eqnarray}
 \widehat{L}_{tb}^{AP}(N) &=& L_{tb}^{P}(N)N^{2(\delta-1)}(1-x_{1.}/N)^{\delta-1}(1-x_0/N)^{1/2}\nonumber\\
 &=& L_{tb}^{MP}(N)N^{2(\delta-1)}(1-x_{1.}/N)^{\delta-1}\nonumber\\
   &=& \frac{N!}{(N-x_0)!}N^{\delta-N-3/2}(N-x_{1.})^{\delta-1}(N-x_0)^{N-x_0+1/2}\nonumber
\end{eqnarray}
\end{center}
\end{result}
\paragraph{}
Now the following theorem justifies the condition on the domain of $\delta$ in order to have a finite maxima for the adjusted profile likelihood for $M_{tb}$. Proof is in the \textit{Appendix}.

\begin{theorem}\label{Theo_5}
(a) Finite maximum adjusted profile likelihood estimate of $N$ exists for the model $M_{tb}$ only if $\delta<1$.\\
(b) For the model $M_{tb}$, $\exists$ some $\delta_0<1 \ni \forall \delta<\delta_0$, $\widehat{L}_{tb}^{AP}(N)\downarrow N$ and hence, corresponding \textit{mle} of $N$ tend to the lower bound ($x_0+1$).
\end{theorem}
\paragraph{}
Hence, a choice of $\delta$, either very small or greater than 1, would lead us to trivial results. Now we try to find a suitable $\delta$ (between $\delta_0$ and $1$), rather a class of suitable $\delta$, in order to obtain a reasonable estimate of $N$. Considering $N$ as real, we found the first derivative of adjusted profile log-likelihood as
$(\partial/\partial N)\widehat{\ell}_{tb}(N)=(\delta-1)/N+(\delta-1)/(N-x_{1.})+A_N$, where sequence $A_N$ is positive and equivalent to $O(N^{-2})$ for fixed data since \textit{digamma function} $\beta(N)=O(N^{-1})$. Equating this to zero we have, $(1-\delta)O(N^{-1})=A_N$ and this implies $\delta=1-B_N$, where $B_N$ is positive sequence of $N$ and equivalent to $O(N^{-1})$.
In practice, one can choose a $\delta$ such that $\delta=1-O_p(N^{-1})$.
\begin{remark}
If we apply the proposed adjustment to the profile likelihood function associated with model $M_t$, then $\widehat{L}_{t}^{AP}(N)$ can be expressed as
\[ \widehat{L}_{t}^{AP}(N)=\mbox{$L_{t}^{MP}(N)N^{2(\delta-1)}$, for all $N\geq x_0$.}\]
\end{remark}
For the model $M_t$, analogous to theorem \ref{Theo_5}, we have the following observations:
\begin{remark}
(a) there exists some $\delta_0<1 \ni \forall \delta<\delta_0$, $\widehat{L}_{t}^{AP}(N)\downarrow N$ and hence, corresponding \textit{mle} of $N$ tend to the lower bound $x_0$,\\
(b) $\exists$ some $\delta'>1 \ni \forall \delta>\delta'$, $\widehat{L}_{t}^{AP}(N)$ does not have finite estimates.
\end{remark}

\subsection{Variance of $\hat{N}_{tb}^{AP}$}\label{subsec_AdPL_var}
\begin{figure}
\begin{minipage}{15.5cm}
  \centering
  \includegraphics[width=6in,height=5in]{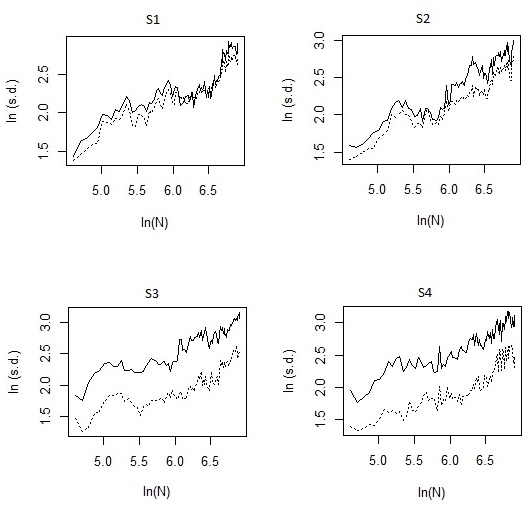}
  \caption{Comparative plots of $log_{e}\{s.d.(\hat{N})\}$ for both estimates $\hat{N}_{tb}^{AP}$ (dotted line) and $\hat{N}_{t}$ (continuous line) over several true $log_{e}(N)$ are plotted for the artificially simulated populations using capture probabilities mentioned in S1, S2, S3 and S4.}
  \end{minipage}
  \label{Fig:1}
\end{figure}

It is found in section \ref{subsec_model_mis} that s.e.($\hat{N}_{t}$) is $O(N^{1/2})$ when independence holds. Hence, to study the nature of variability in $\hat{N}_{tb}^{AP}$, can we postulate that $s.e.(\hat{N}_{tb}^{AP})=O(N^{\alpha})$, for some $\alpha>0$?
To investigate this and if so, to get some idea on the extent of $\alpha$, we take the following example. Finally, a comparison of the pattern of variability in $\hat{N}_{tb}^{AP}$ against $\hat{N}_{t}$ under the underlying model $M_{tb}$ is made graphically.
\paragraph{}
\textit{Example:} Let us consider four artificial populations S1($p_{1.}=0.60$, $p_{.1}=0.70$), S2($p_{1.}=0.70$, $p_{.1}=0.55$), S3($p_{1.}=0.60$, $p_{.1}=0.70$) and S4($p_{1.}=0.70$, $p_{.1}=0.55$) following model $M_{tb}$. From each population, we generate 200 data sets ($x_{11},x_{.1},x_{1.}$) and obtain $\hat{N}_{tb}^{AP}$ for each data. These 200 estimates constitutes the sampling distributions of the estimator. Finally, s.d. over 200 replicates is calculated to measure s.e. of the estimate. Same calculations are also done for the estimator $\hat{N}_{ind}=(x_{1.}x_{.1}/x_{11})$ and finally, comparative behaviour of the $ln(s.e.)$ of both the estimators $\hat{N}_{ind}$ and $\hat{N}_{tb}^{AP}$ are plotted against $ln(N)$ in Figure \ref{Fig:1}. Figure shows that s.e.($\hat{N}_{tb}^{AP}$) is less than s.e.($\hat{N}_{t}$) $\forall N$ and values of estimated $\alpha$ in s.e.($\hat{N}_{tb}^{AP}$) are between $0.25$ and $0.30$ for all the populations.
Thus, the numerical investigations carried out above suggests that the proposed adjusted profile likelihood could be more helpful in the context of population size estimation (under the model $M_{tb}$) and it shows better efficiency than the usual DSE estimator $\hat{N}_{ind}$ in terms of s.e.

\section{Numerical Illustrations}\label{Sec:NumericalIllustrations}
\subsection{Simulation Study 1}\label{subsec:Simulation1}
In this section we have considered various artificial populations, reflecting different possible situations under $M_{tb}$, to illustrate the behaviour of the competitive estimators in DRS discussed in earlier sections under the model $M_{tb}$. In any kind of time ordered samples, the possible list-dependence can be modelled through $M_{tb}$. First, we simulated four populations for each behavioral dependence situation ($\phi=0.80$ and $\phi=1.25$ respectively represents the recapture averseness and recapture proneness) that encompasses all possible combinations. Capture probabilities for those populations, each having size $N=500$, are structurally presented in Table \ref{Tab:2}. The expected number of distinct captured individuals ($E(x_0)=N(p_{11}+p_{01}+p_{10})$) for each population is cited in Table \ref{Tab:2}.
\begin{table}[ht]
\centering
\caption{Populations with $N=500$ considered for simulations study}
\begin{minipage}{15cm}
\begin{tabular}{|ccccccccccc|}
\hline
Population & $\phi$ & $p_{1.}$ & $p_{.1}$&  $E(x_0)$ & & Population & $\phi$ & $p_{1.}$ & $p_{.1}$ & $E(x_0)$\\
\hline
\hline
P1 & 1.25 & 0.50 & 0.65 & 394  & &  P5 & 0.80 & 0.50 & 0.65 & 430\\
P2 & 1.25 & 0.60 & 0.70 & 422  & & P6 & 0.80 & 0.60 & 0.70 & 459\\
P3 & 1.25 & 0.80 & 0.70 & 458  & & P7 & 0.80 & 0.80 & 0.70 & 483\\
P4 & 1.25 & 0.70 & 0.55 & 420  & & P8 & 0.80 & 0.70 & 0.55 & 446\\
\hline
\end{tabular}
\end{minipage}
\label{Tab:2}
\end{table}
It is noted that in the first two populations for each $\phi$, $p_{1.}<p_{.1}$ which refers to the usual situation in DRS data obtained by a specialised survey conducted after a large census operation, e.g. Post Enumeration Survey (PES). The last two populations with $p_{1.}>p_{.1}$ are just the opposite case which is observed often in a study of the estimation of drug users.
It is also noted that P2, P4, P6 and P8 are same as hypothetical populations S1, S2, S3 and S4 respectively, considered for illustration of the variance of proposed estimates in section \ref{subsec_AdPL_var}.
Now, 200 data sets ($x_{11}, x_{.1}, x_{1.}$) are generated from each of the above eight populations. We present the adjusted profile likelihood estimate (AdPL) for each situations for different reasonable $\delta$ values. To compare the performance of our proposed method with Bayesian strategy, we compute the estimates by Lee et al. (2003 \cite{Lee03}). In addition, the estimates assuming list-independence, $\hat{N}_{ind}$, are also shown to empirically understand the extent of bias due to model mis-specification discussed in section \ref{subsec_model_mis}. For each estimate, several other frequentist measures are shown to evaluate the relative performance of the said estimators.
Final estimates of \emph{N} is obtained by averaging over 200 replications. Based on those 200 estimates, the sample s.e., sample RMSE (Root Mean Square Error) and $95\%$ bootstrap confidence interval (C.I.) are also presented in Table \ref{Tab:3} (for $\phi=1.25$ representing \textit{recapture-prone} situations) and Table \ref{Tab:4} (for $\phi=0.80$ representing \textit{recapture-averse} situations). For Lee's Bayes estimates, $95\%$ credible interval (C.I.) based on sample quantile of the marginal posterior distribution of $N$ is presented.
\begin{table}[ht]
\centering
\caption{Summary results for populations P1-P4 (representing \textit{recapture-prone} situations) when \textbf{No} directional information on $\phi$ is available.}
\begin{minipage}{16.5cm}
\begin{tabular}{|lcrcccc|}
\hline
Method & & & P1 & P2 & P3 & P4\\
\hline
\hline
& & & &  &  & \\
$\hat{N}_{ind}$& & $\hat{N}$(s.e.) & 450(14.10) & 460(11.23) & 480(7.07) & 469(12.01)\\
 & & RMSE  & 51.54 & 41.32 & 20.55 & 32.55\\
 &  & C.I.  & $(425, 480)$ & $(438, 481)$ & $(465, 493)$ & $(444, 491)$\\
\multicolumn{7}{|c|}{}\\
Lee\footnote{with prior $\pi(\phi)=$ U($0.5,2$). [Chatterjee and Mukherjee, p.p. 14 (2014 \cite{Chatterjee14})]}& & $\hat{N}$(s.e.) & 468(20.56) & 483(18.45) & 485(6.61) & 471(8.11)\\
 & & RMSE  & 37.94 & 24.97 & 16.97 & 30.61\\
 &  & C.I.  & $(398, 561)$ & $(426, 560)$ & $(460, 513)$ & $(422, 542)$\\
\multicolumn{7}{|c|}{}\\
AdPl & $\delta=1-0.75N^{-1}$ & $\hat{N}$(s.e.) & 486(12.15) & 513(10.61) & 539(7.15) & 499(9.74)\\
 & & RMSE  & 18.86 & 17.01 & 39.82 & 9.61\\
 &  & C.I.  & $(461, 507)$ & $(491, 532)$ & $(525, 552)$ & $(578, 516)$\\
\multicolumn{7}{|c|}{}\\
 & $\delta=1-1.25N^{-1}$ & $\hat{N}$(s.e.) & 461(11.47) & 488(10.01) & 515(6.78) & 476(9.27)\\
 & & RMSE  & 40.32 & 15.54 & 16.32 & 25.68\\
 &  & C.I.  & $(439, 480)$ & $(467, 506)$ & $(501, 527)$ & $(456, 493)$\\
\multicolumn{7}{|c|}{}\\
 & $\delta=1-1.75N^{-1}$ & $\hat{N}$(s.e.) & 449(11.13) & 476(9.77) & 504(6.60) & 466(9.02)\\
 & & RMSE  & 51.64 & 25.85 & 7.71 & 35.23\\
 &  & C.I.  & $(428, 469)$ & $(455, 493)$ & $(491, 516)$ & $(446, 482)$\\
\hline
\end{tabular}
\end{minipage}
\label{Tab:3}
\end{table}
\begin{table}[ht]
\centering
\caption{Summary results for populations P5-P8 (representing \textit{recapture-averse} situations) when \textbf{No} directional information on $\phi$ is available.}
\begin{minipage}{16.5cm}
\begin{tabular}{|lcrcccc|}
\hline
Method & & & P5 & P6 & P7 & P8\\
\hline
\hline
& & & &  &  & \\
$\hat{N}_{ind}$ & & $\hat{N}$(s.e.) & 563(23.15) & 550(14.94) & 526(8.08) & 538(14.26)\\
 & & RMSE  & 67.21 & 52.48 & 27.09 & 40.44\\
 & & C.I.  & $(523, 615)$ & $(524, 578)$ & $(510, 541)$ & $(513, 565)$\\
\multicolumn{7}{|c|}{}\\
Lee\footnote{with prior $\pi(\phi)=$ U($0.5,2$). [Chatterjee and Mukherjee (2014, p.p. 18 \cite{Chatterjee14})]} & & $\hat{N}$(s.e.) & 474(20.80) & 512(15.76) & 516(6.17) & 517(13.02)\\
& & RMSE  & 35.58 & 19.83 & 18.71 & 21.75\\
&  & C.I.  & $(431, 566)$ & $(461, 575)$ & $(486,553)$ & $(451, 615)$\\
\multicolumn{7}{|c|}{}\\
AdPl & $\delta=1-0.75N^{-1}$ & $\hat{N}$(s.e.) & 533(9.53) & 562(7.44) & 574(5.70) & 536(8.15)\\
 & & RMSE  & 34.57 & 63.05 & 74.25 & 36.88\\
 &  & C.I.  & $(513, 552)$ & $(547, 577)$ & $(563, 584)$ & $(521, 551)$\\
\multicolumn{7}{|c|}{}\\
 & $\delta=1-1.25N^{-1}$ & $\hat{N}$(s.e.) & 505(9.40) & 534(6.98) & 548(5.21) & 510(7.75)\\
 & & RMSE  & 10.72 & 35.23 & 48.40 & 13.01\\
 &  & C.I.  & $(487, 524)$ & $(519, 547)$ & $(537, 557)$ & $(497, 525)$\\
\multicolumn{7}{|c|}{}\\
 & $\delta=1-1.75N^{-1}$ & $\hat{N}$(s.e.) & 492(9.18) & 521(6.75) & 535(5.00) & 499(7.52)\\
 & & RMSE  & 12.45 & 22.04 & 35.88 & 9.65\\
 &  & C.I.  & $(474, 510)$ & $(506, 534)$ & $(525, 545)$ & $(485, 512)$\\
\hline
\end{tabular}
\end{minipage}
\label{Tab:4}
\end{table}
\paragraph{}
Table \ref{Tab:3} says that as $\delta(<1)$ is chosen to be closer to 1, AdPL performs better for case of low capture probabilities (P1 \& P4). In other situations (P2 \& P3) where capture probabilities are high, efficient adjustment coefficient $\delta$ will be ($1-1.25N^{-1}$). In other words, we try to analyse the performance from the perspective of two kinds of populations where $x_{1.}<x_{.1}$ and $x_{1.}>x_{.1}$. For both kind of situations $x_{1.}<x_{.1}$ (\textit{i.e.} P1 \& P2) and $x_{1.}>_{10}$ (\textit{i.e.} P3 \& P4), AdPL performs progressively better as $\delta(<1)$ is chosen to be closer to 1. Except P3, AdPL shows more efficient result than Lee's method. In any recapture prone situation, the use of $\hat{N}_{ind}$ will certainly mislead us, particularly for the cases where capture probabilities are low and/or when underlying $\phi$ is far above 1.
\paragraph{}
Similarly, when we turn to analyse some considered hypothetical populations with recapture averseness, we see from Table \ref{Tab:4} that as $\delta$ is chosen to be relatively smaller at ($1-1.75N^{-1}$), AdPL performs reasonably better. In low capture situations (P5 and P8), AdPL shows more efficient result than Lee's method. Table \ref{Tab:4} also shows that in any recapture averse situations, $\hat{N}_{ind}$ will highly overestimate $N$ as $\phi$ is substantially different from 1.
\paragraph{}
Hence, in both situations of recapture aversion and proneness, poor performance of $\hat{N}_{ind}$ becomes worse particularly for the populations where $x_{1.}<x_{.1}$. Lee's Bayes estimate, with prior $\pi(\phi)=U(0.5,2)$, generally underestimates for $\phi>1$ and overestimates for $\phi<1$ but use of their estimate is recommended than that of $\hat{N}_{ind}$ to avoid serious model mis-specification. However, we found that our proposed adjusted profile likelihood method, with suitably chosen value of $\delta$, can perform better than Lee's.

\subsection{Simulation Study 2}\label{subsec:Simulation2}
Here we examine some frequentist as well as robustness properties of the adjusted profile-likelihood estimate along with the simple estimate $\hat{N}_{ind}=$($x_{1.}x_{.1}/x_{11}$).

\subsubsection*{\textit{Frequentist Coverage Performance:}}\label{subsubsec_frequentist_coverage}
\begin{figure}
  \centering
  \includegraphics[width=6in,height=5in]{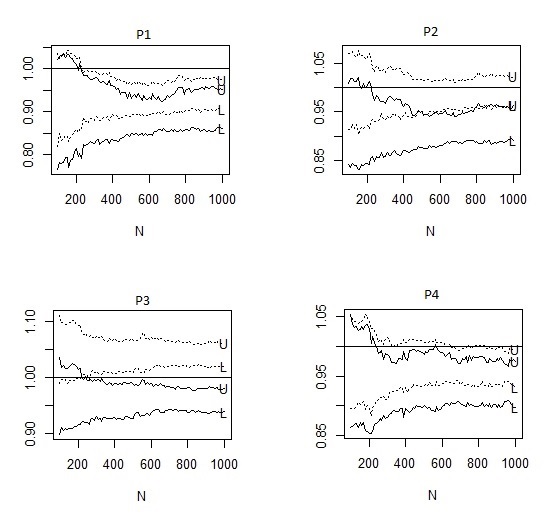}
  \caption{Comparative plots of confidence bands of $\hat{N}/N$ corresponding to both the estimates $\hat{N}_{tb}^{AP}$ (dotted line) and $\hat{N}_{t}$ (continuous line) are plotted against different true $N$ for populations P1-P4 (recapture-prone cases). The targeted value of $\hat{N}/N$ is indicated at 1.0 (presenting unbiasedness).}
  \label{Fig:2}
\end{figure}

\begin{figure}
  \centering
  \includegraphics[width=6in,height=5in]{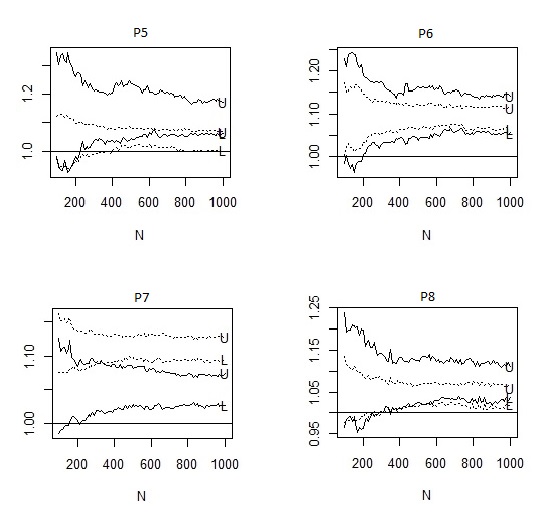}
  \caption{Comparative plots of confidence bands of $\hat{N}/N$ corresponding to both the estimates $\hat{N}_{tb}^{AP}$ (dotted line) and $\hat{N}_{t}$ (continuous line) are plotted against different true $N$ for populations P5-P8 (recapture-averse cases). The targeted value of $\hat{N}/N$ is indicated at 1.0 (presenting unbiasedness).}
  \label{Fig:3}
\end{figure}

Firstly, under the mis-specification threat (\textit{see} section \ref{subsec_model_mis}), we graphically study the coverage performance of $\hat{N}_{ind}=$($x_{1.}x_{.1}/x_{11}$) for true $N$ as $N$ varies. Moreover to compare with the $\hat{N}_{tb}^{AP}$, we also do same for our proposed AdPL estimator. We consider all the artificial populations (following $M_{tb}$) simulated earlier in section \ref{subsec:Simulation1}. For moderately large population (say, $N>100$), we found both the $\hat{N}_{ind}$ and $\hat{N}_{tb}^{AP}$ to be approximately normal. Figure \ref{Fig:2} and \ref{Fig:3} show simultaneous plot of the $95\%$ \textit{relative UCL} (=$(\hat{N}+1.96 s.e.(\hat{N}))/N$) and \textit{LCL} (=$(\hat{N}-1.96 s.e.(\hat{N}))/N$) corresponding to the estimators $\hat{N}_{ind}$ and $\hat{N}_{tb}^{AP}$ over several true $N$. The motivation behind this unorthodox type of figures is as follows. The \textit{Relative LCL} and \textit{relative UCL} contains 1 with $0.95$ probability. Hence, we can compare how much the relative confidence limits for the said estimators deviate from 1 with gradually increasing true $N$ (here, it ranges from 100 to 1000).
For the recapture prone ($\phi>1$) cases, Figure \ref{Fig:2} shows that relative confidence bounds of $\hat{N}_{tb}$ are slightly tighter as well as closer to 1 in most of the situations compared to $\hat{N}_{ind}$. Analogously, Figure \ref{Fig:3}, for the recapture aversion ($\phi<1$) cases, shows that confidence bounds of $\hat{N}_{tb}$ are tighter than that of $\hat{N}_{ind}$ as $N$ increases and it is relatively closer to 1 in all situations for different $N$ values.

\subsubsection*{\textit{Robustness Consideration:}}\label{subsubsec_robustness}
\begin{figure}
  \centering
  \includegraphics[width=6.5in,height=5in]{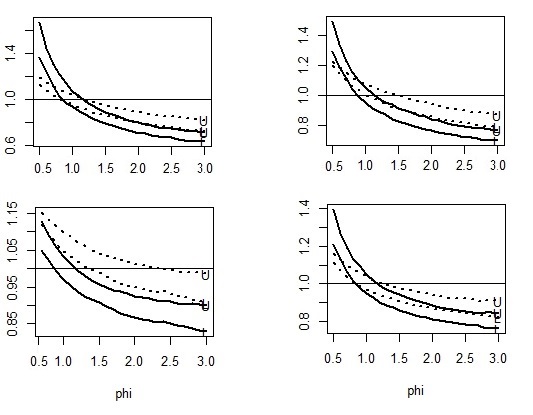}
  \caption{Comparative plots of confidence bands of $\hat{N}/N$ corresponding to both the estimates $\hat{N}_{tb}^{AP}$ (dotted line) and $\hat{N}_{t}$ (continuous line) are plotted against different $\phi$ for four situations. The targeted value of $\hat{N}/N$ is indicated at 1.0 (presenting unbiasedness).}
  \label{Fig:4}
\end{figure}

Our other interest is on the robustness of the proposed estimator and the usual C-D estimator $\hat{N}_{ind}$. Actually the model $M_{tb}$ is driven by the unidentifiable behavioral effect parameter $\phi$.
An useful estimator for $N$ should be as robust as possible with respect to the underlying $\phi$ value and hence, in Figure \ref{Fig:4}, we present a comparative study on both the estimates against different $\phi$. We fix true $N$ at 500 and $\phi$ is considered to vary between $0.5$ and $3.0$. In simulation 1, four artificial situations are assumed without considering the $\phi$ value. Here we have studied the
robustness for all those four situations.
Figure \ref{Fig:4} depicts that $\hat{N}_{tb}^{AP}$ has better robustness w.r.t. $\phi$ than $\hat{N}_{ind}$ in all situations.

\subsection{Real Data Illustrations}\label{subsec:RealData}
\subsubsection{Example 1}\label{subsubsec:RealData01}
An example of DRS data is considered on death count obtained from a Population Change Survey conducted by the National Statistical Office in Malawi between 1970 and 1972 (for details, \textit{see} Greenfield (1975 \cite{Green75}). Only two strata, called Lilongwe ($\hat{c}=0.593$, $x_{.1}>x_{1.}$) and Other urban areas ($\hat{c}=0.839$, $x_{.1}<x_{1.}$), are selected to illustrate the role of different $\hat{c}$ values and opposite nature of $x_{.1}$ and $x_{1.}$. Significantly lower $\hat{c}$ value indicates that the people of Lilongwe seemed to be less likely to give the information on deaths again in survey time than that of \emph{Other urban areas} people.
\paragraph{}
Now, if anyone wishes to use the widely acceptable model $M_t$ assuming list-independence and calculate the simple estimate $\hat{N}_{ind}$, he/She would find that 365 and 2920 deaths occurred in \textit{Lilongwe} and \textit{Other urban areas} respectively. Nour (1982 \cite{Nour82}) argued that the assumption of independent collection procedures is unacceptable in reality. Assuming the fact that two data sources are positively correlated (i.e. $\phi>1$) in a human demographic study, they estimated death sizes as \textit{378} (i.e. $\hat{\phi}=1.33$) and \textit{3046} (i.e. $\hat{\phi}=1.13$) for \textit{Lilongwe} and \textit{Other urban areas} respectively. However, in this article we do not make any such assumptions on the directional nature of $\phi$. We consider the data as just an $2\times2$ DRS data where nothing is known about $\phi$. Then, Lee et al.'s fully Bayes method with uniform prior $\pi(\phi)=U(0.1,2)$ finds that 372($\hat{\phi}=1.19$) and 3205($\hat{\phi}=1.30$) deaths occured in \textit{Lilongwe} and \textit{Other urban areas} respectively. Our adjusted profile likelihood method estimates the
death sizes as 378($\hat{\phi}=1.33$) and 3428($\hat{\phi}=1.53$) respectively, taking $\delta=1-4(1-\hat{c})N^{-1}$. Our estimates agree with Nour's for \textit{Lilongwe} but Nour's estimate for \textit{Other urban areas} is significantly smaller than Lee's estimate as well as our estimate.
\subsubsection{Example 2}\label{subsubsec:RealData02}
Another example of DRS data is considered on injection drug user (IDU) of greater Victoria, British Columbia,
Canada (Xu et al., 2014 \cite{Xu14}). To track the changes in the prevalence of HIV and hepatitis C, the Public Health Agency of Canada developed the national, cross-sectional I-Track survey. Phase I and phase II of the I-Track survey were completed in Victoria in 2003 and 2005, respectively. With only two samples from the I-Track survey (phase I and phase II), some closed population mark-recapture models were implemented to estimate the number of IDUs in greater Victoria, BC. They found that Lincoln-Petersen (LP) estimate, $\hat{N}_{ind}$ from model $M_t$, for the total number of injection drug users was 3329. They also commented that LP estimator might not be worthwhile if independent assumption was violated when \textit{behaviour response} and/or \textit{heterogeneity} affects the probability of capture. They use Huggins' (1989 \cite{Hugg89}) conditional likelihood approach to deal with plausible heterogeneity in the data and estimate was 3342. Moreover, the time ordering of samples offers an opportunity to use model $M_{tb}$. Literature on epidemiological studies on such type of ‘hidden’ or ‘hard to reach’ population says that individual, who are listed in first survey, tries to avoid the listing operation in second survey. Thus there is high possibility of \textit{recapture-aversion} (i.e. $\phi<1$). Low recapture rate is ($\hat{c}=0.075$), which strengthens this possibility.
\paragraph{}
Considering the DRS data originated from model $M_{tb}$ with $\phi>0$, Lee et al.'s fully Bayes method with prior $\pi(\phi)=U(0.01,2)$ finds that 596($\hat{\phi}=0.11$) number of drug users are in that population. As $\hat{c}$ is found very low, our adjusted profile likelihood method estimates the size of injection drug users as 584 ($\hat{\phi}=0.09$) taking $\delta=1-4(1-\hat{c})N^{-1}$. Hence, Lee's method and our adjusted profile likelihood method says that if you consider the population as quite homogeneous then most general model $M_{tb}$ suggests that total number of injection drug user of greater Victoria is around 580 to 600, a much lower estimate than the estimate of drug users under independence.

\section{Summary and Conclusions}\label{sec_Conclusion}
In the context of population size ($N$) estimation, inappropriateness of model $M_t$ is advocated for Dual-record system (DRS) in several real life situations. But at present this model is widely employed specially in census undercount estimation and epidemiology due to its simplicity. We have considered the most general model $M_{tb}$ that allows the behaviour response effect to play a significant role along with time variation effect in estimating $N$. The model $M_{tb}$ suffers from identifiability problem where suitable Bayesian methods might have the potential to overcome that burden. However, in this article we have investigated the usefulness of pseudo likelihood approaches based on profiling the interest parameter $N$. Ordinary profile, modified profile and approximated modified profile likelihoods have been shown to be useless for model $M_{tb}$. An adjustment on profile likelihood (AdPL) is proposed tuned by an adjustment coefficient so that reasonably better solution can be made available. The present article also shows mathematical and graphical analyses of possible model mis-specification due to the use of $M_t$.
\paragraph{}
The proposed method depends on the choice of $\delta$ (close to $1-N^{-1}$) using the knowledge of $\hat{c}$ and possible direction of $\phi$. In real life situations, if $\phi$ is unknown, then uniform choice is possible. Lee et al. (2003 \cite{Lee03}) Bayes method provides better coverage than any other method but also it possesses lower efficiency in most situations than AdPL. Moreover, Lee's method, with trial-and-error approach to discover a suitable range for uniform prior $\pi(\phi)$, may take a long time. Some other disadvantages are subjectiveness of the informative prior $\pi(\phi)$, highly dispersed conditional posterior of $\phi$, etc. Thus, our proposed
adjusted method is useful to obtain an efficient estimate of population size ($N$)
very quickly from this complex DRS. In addition to that, AdPL helps to produce more efficient alternatives specially in recapture prone situations.


\section*{Appendix}
\textit{Proof of Theorem \ref{Theo_1}:}\\
At first we shall derive the Bias of $(x_{1.}x_{.1}/x_{11})$ in terms of original DRS probabilities in Table \ref{tab:dse}. In multinomial setup, we have $E(x_{ab})=Np_{ab},Cov(x_{ab},x_{cd})=-Np_{ab}p_{cd}$, for $a,b,c,d\in\{1,2\}$. Then replacing $x$, $y$ and $z$ by $x_{1.}$, $x_{.1}$ and $x_{11}$ respectively in the above $Lemma 1$, we have
\begin{center}
\begin{eqnarray}
E(x_{1.}x_{.1}/x_{11})&=&Np_0+\frac{Np_{01}p_{10}}{p_{11}}\left(1+\frac{1}{N}+\frac{1-p_{11}}{Np_{11}}\right)\nonumber\\
&=&Np_0+\frac{Np_{01}p_{10}}{p_{11}}\left(1+\frac{1}{Np_{11}}\right)\nonumber\\
&=&Np_0+\frac{Np_{01}p_{10}}{p_{11}}+\frac{p_{01}p_{10}}{p_{11}^2}\nonumber
\end{eqnarray}
\end{center}
Hence, Bias($x_{1.}x_{.1}/x_{11}$)=$E(x_{1.}x_{.1}/x_{11})-N=-N(1-p_0)+N(p_{01}p_{10}/p_{11})+(p_{01}p_{10}/p_{11}^2)$. Now, in $M_{tb}$, $c=\phi p=p_{11}/p_{1.}$ and $p=p_{01}/(1-p_{1.})$. Hence, after some algebraic simplification, we found Bias($x_{1.}x_{.1}/x_{11}$)=$N(1-p_{1.})(1-\phi)/\phi+\frac{(1-p_{1.})(1-\phi p)}{p_{1.}\phi^2 p}$. $\qed$
\\
\\
\textit{Proof of Theorem \ref{Theo_2}:}\\
Atfirst define $R^{P}_{t}(N)=L_{t}^{P}(N+1)/L_{t}^{P}(N)$ and after some algebraic simplification we have, $R^{P}_{t}(N)=\frac{(N-x_{1.}+1)(N-x_{.1}+1)}{(N-x_{0}+1)(N+1)}(\frac{N}{N+1})^{2N}(1+\frac{1}{N-x_{1.}})^{N-x_{1.}}(1+\frac{1}{N-x_{.1}})^{N-x_{.1}}$.
Now, $\frac{(N-x_{1.}+1)(N-x_{.1}+1)}{(N-x_{0}+1)(N+1)}\times$\\$(\frac{N}{N+1})^{2N}\geq1\Rightarrow \frac{(N-x_{1.}+1)(N-x_{.1}+1)}{(N-x_{0}+1)(N+1)}>1$, and that holds for all $N<(x_{1.}x_{.1}/x_{11})-1$.
Therefore, $R^{P}_{t}(N)>1$ for all $N<(x_{1.}x_{.1}/x_{11})-1$. Hence, corresponding \textit{mle} $\hat{N}_{t}^{P}$ is $(x_{1.}x_{.1}/x_{11})-1$ when $(x_{1.}x_{.1}/x_{11})$ is integer. When, $(x_{1.}x_{.1}/x_{11})$ is not an integer, $\hat{N}_{t}^{P}$ equal to either $[x_{1.}x_{.1}/x_{11}]-1$ or $[x_{1.}x_{.1}/x_{11}]$,
which attains the maximum value of the profile likelihood $L_{t}^{P}(N)$, where $[u]$ denotes the greatest integer less than or equal to $u$, for $u\in\mathbb{R}$. Thus, in general, $\hat{N}_{t}^{P}=[x_{1.}x_{.1}/x_{11}]-1$ or $[x_{1.}x_{.1}/x_{11}]$ and $\hat{N}_{t}^{P}$ is finite iff $x_{11}>0$. $\qed$
\\
\\
\textit{Proof of Result \ref{Res_1}:}\\
According to parametrization $\theta=N$ and $\psi$=($p_{1.},p_{.1}$), it is straightforward to show that the log-likelihood for model $M_t$,\\
\[\ell^{t}(\theta,\psi)=\sum_{i=1}^{x_0}ln(N-x_0+i)+x_{1.}lnp_{1.}+x_{.1}lnp_{.1}+(N-x_{1.})ln(1-p_{1.})+(N-x_{.1})ln(1-p_{.1}).\]
Hence, $\ell_{\psi}^{t}(\theta,\psi)=\left(\frac{x_{1.}}{p_{1.}}-\frac{N-x_{1.}}{1-p_{1.}},\frac{x_{.1}}{p_{.1}}-\frac{N-x_{.1}}{1-p_{.1}}\right)$ and
\[E\{\ell_{\psi}^{t}(\theta,\psi);\theta_0,\psi_0\}\left|_{\theta_0=\hat{\theta},\psi_0=\hat{\psi}}\right.=\left(\frac{\hat{N}\hat{p}_{01}}{p_{1.}}-\frac{N-\hat{N}\hat{p}_{01}}{1-p_{1.}},\frac{\hat{N}\hat{p}_{10}}{p_{.1}}-\frac{N-\hat{N}\hat{p}_{10}}{1-p_{.1}}\right).\]
Therefore, $|I^{t}(\theta,\hat{\psi}_{\theta};\hat{\theta},\hat{\psi})|=\frac{N^4}{(N-x_{1.})(N-x_{.1})}$, since $\hat{\psi}_{\theta}=\left(\frac{x_{1.}}{N},\frac{x_{.1}}{N}\right)$ and
\begin{center}
\begin{eqnarray}
I^{t}(\theta,\psi;\hat{\theta},\hat{\psi})&=&\frac{\partial}{\partial\hat{\psi}} E\{\ell_{\psi}^{t}(\theta,\psi);\theta_0,\psi_0\}\left|_{\theta_0=\hat{\theta},\psi_0=\hat{\psi}}\right.\nonumber\\
&=&\left(\begin{tabular}{cc}
         $\frac{\hat{N}}{p_{1.}}+\frac{\hat{N}}{1-p_{1.}}$ & 0\nonumber\\
         0 & $\frac{\hat{N}}{p_{.1}}+\frac{\hat{N}}{1-p_{.1}}$\nonumber\\
         \end{tabular}
   \right)
\end{eqnarray}
\end{center}
Again, from Severini (2000), we have $|\ell_{\psi;\hat{\psi}}^{t}(\theta,\hat{\psi}_{\theta})|=\frac{N^4}{(N-x_{1.})(N-x_{.1})}$, ignoring the terms not depending on data. So,
it is clear that $|\ell_{\psi;\hat{\psi}}^{t}(\theta,\hat{\psi}_{\theta})|=|I^{t}(\theta,\hat{\psi}_{\theta};\hat{\theta},\hat{\psi})|$. Thus, from \textit{remark \ref{rem_1}}, $L^{MP}(\theta)=\widetilde{L}^{MP}(\theta)$ for $M_{t}$ and
$\hat{j}_{\psi\psi}^{t}(\theta,\hat{\psi}_{\theta})=-\ell_{\psi;\psi}(\theta,\hat{\psi}_{\theta})=Diag\{\frac{N^3}{N-x_{1.}},\frac{N^3}{N-x_{.1}}\},$ which leads to the proof of this result using (\ref{Eq_MPL_2}). $\qed$
\\
\\
\textit{Proof of Theorem \ref{Theo_3}:}\\
Let us define $R^{MP}_{t}(N)=L_{t}^{MP}(N+1)/L_{t}^{MP}(N)$. Then we have $R^{MP}_{t}(N)=R^{P}_{t}(N)\times$\\$\frac{(N-x_{1.}+1)^{1/2}(N-x_{.1}+1)^{1/2}}{(N-x_{1.})^{1/2}(N-x_{.1})^{1/2}}\frac{N}{(N+1)}$, where $R^{P}_{t}(N)=L_{t}^{P}(N+1)/L_{t}^{P}(N)$.
Now, by some algebraic manipulation it can be shown that $\frac{(N-x_{1.}+1)^{1/2}(N-x_{.1}+1)^{1/2}}{(N-x_{1.})^{1/2}(N-x_{.1})^{1/2}}\frac{N}{(N+1)}\geq1$ for all $N\geq\frac{2x_{1.}x_{.1}}{(x_{1.}+x_{.1})}$. Moreover, $\frac{2x_{1.}x_{.1}}{(x_{1.}+x_{.1})}<x_0$ always.
So, $R^{MP}_{t}(N)\geq R^{P}_{t}(N)>1$ for all $x_0\leq N<(x_{1.}x_{.1}/x_{11})-1$. Therefore, the maximum profile likelihood estimate $\hat{N}_{t}^{P}$ is always less than or equal to the the maximum modified profile likelihood estimate $\hat{N}_{t}^{MP}$. $\qed$
\\
\\
\textit{Proof of Theorem \ref{Theo_4}:}\\
From \textit{Theorem} \ref{Theo_1} and \ref{Theo_2} we have $R^{MP}_{t}(N)\geq R^{P}_{t}(N)>1$ for all $N<(x_{1.}x_{.1}/x_{11})-1$. Now, if $L_{t}^{P}(N)$ is maximum at $N=\widetilde{N}$ (say), then $R^{P}_{t}(\widetilde{N})\leq1<R^{P}_{t}(\widetilde{N}-1)\leq R^{MP}_{t}(\widetilde{N}-1)$ if $\widetilde{N}-1\geq x_0$.
Since $R^{P}_{t}(\widetilde{N})\leq R^{MP}_{t}(\widetilde{N})$ for $\widetilde{N}\geq x_0$ i.e. $(x_{10}x_{01}/x_{11})>1$, one have to check whether $R^{MP}_{t}(\widetilde{N})>1$ or not, for different possible $\widetilde{N}$.

Now, it is clear that if $\widetilde{N}=[x_{1.}x_{.1}/x_{11}]-1$, $R^{MP}_{t}(\widetilde{N})>1$ since $[x_{1.}x_{.1}/x_{11}]-1\leq(x_{1.}x_{.1}/x_{11})-1$, therefore $\hat{N}_{t}^{MP}=[x_{1.}x_{.1}/x_{11}]$.

If $\widetilde{N}=[x_{1.}x_{.1}/x_{11}]$, $R^{MP}_{t}(\widetilde{N})<1$ since $[x_{1.}x_{.1}/x_{11}]>(x_{1.}x_{.1}/x_{11})-1$, therefore $\hat{N}_{t}^{MP}=[x_{1.}x_{.1}/x_{11}]$.

When $(x_{1.}x_{.1}/x_{11})$ is integer, $\widetilde{N}=(x_{1.}x_{.1}/x_{11})-1$, therefore $R^{MP}_{t}(\widetilde{N})<1$, hence $\hat{N}_{t}^{MP}=(x_{1.}x_{.1}/x_{11})-1$.

Hence, associated \textit{mle} $\hat{N}_{t}^{MP}$ is equal to $(x_{1.}x_{.1}/x_{11})-1$ if $(x_{1.}x_{.1}/x_{11})$ is an integer; otherwise $\hat{N}_{t}^{MP}=[x_{1.}x_{.1}/x_{11}]$. All estimates are finite iff $x_{11}>0$. $\qed$
\\
\\
\textit{Proof of Theorem \ref{Theo_5}(a):}\\
Let us define $(\partial/\partial N)log\widehat{L}^{AP}(N)=\widehat{\ell}'(N)$. We have $\widehat{\ell}_{tb}^{'}(N)=\beta(N+1)-\beta(N-x_0+1)-logN+(\delta-3/2-N)/N+(\delta-1)/(N-x_{1.})+log(N-x_{0})+(N-x_0+1/2)/(N-x_0)$. After
some algebraic simplification using the asymptotic approximation of \textit{digamma function} $\beta(N)=O(N^{-1})$ we have, $\widehat{\ell}_{tb}^{'}(N)=(\delta-1)/N+(\delta-1)/(N-x_{1.})+A_N$, where $A_N$ is positive quantity decreases to zero and equivalent to $O(N^{-2})$, because $\beta^{'}(N)=O(N^{-2})$.
Clearly, if $\delta=1$, $\widehat{\ell}_{tb}^{'}(N)>0$, for all $N>x_0$. When $\delta>1$, $\widehat{\ell}_{tb}^{'}(N)=O(N^{-1})>0$, for all $N>x_0$.
Therefore, $\widehat{L}^{AP}(N)$ is strictly increasing for $N>x_0$ if $\delta\geq1$ and hence, finite \textit{mle}, $\hat{N}_{tb}^{AP}$,
does not exist for $\delta\geq1$. Again if $\delta<1$, then $\widehat{\ell}_{tb}^{'}(N)=A_N+B_N$, where $B_N=(\delta-1)(2N-x_{1.})/N(N-x_{1.})<0$ is increases to zero. So, there may exist some $N$, for which $\widehat{\ell}_{tb}(N)$ has maxima. If $B_N$ dominates $A_N$ for all $N$, then maxima coincides with the lowest value, i.e. ($x_0+1$). Hence we can certainly establish that, for any $\delta<1$, $(x_0+1)\leq\hat{N}_{tb}^{AP}<\infty$. Thus, finite \textit{mle} for $M_{tb}$ exists only when $\delta<1$. $\qed$
\\
\\
\textit{Proof of Theorem \ref{Theo_5}(b):}\\
In case of model $M_{tb}$, as $L_{tb}^{P}(N)\downarrow N$ for $N\geq x_0$ and $L_{tb}^{MP}(N)\uparrow N$ for $N>x_0$, then from result \ref{Res_2}, we can say that $(1-x_0/N)^{1/2}$ increases in $N$ with a greater rate than the rate of
decrement of $L_{tb}^{P}(N)$. Now, $N^{2(\delta-1)}(1-x_{1.}/N)^{\delta-1}$ decreases with $N$ for $\delta<1$. Therefore, from result \ref{Res_3} one can definitely say that
there must exist some $\delta_0<1 \ni \forall \delta<\delta_0$, $\widehat{L}_{tb}^{AP}(N)\downarrow N$ and hence the proof. $\qed$


\begin{thebibliography}{50}
\bibitem{Barndorff-nielsen83} Barndorff-Nielsen, O. E. (1983), \textit{On a formula for the distribution of the maximum likelihood estimator}, Biometrika \textbf{70}, 343-365.
\bibitem{Barndorff-nielsen85} Barndorff-Nielsen, O. E. (1985), \textit{Properties of modified profile likelihood. In Contributions to Probability and Statistics in Honour of Gunnar Blom, Ed. J. Lankc and G. Lindgren}, Lund: Dept Math. Statist, Lund University., pp. 25-38.
\bibitem{Basu77} Basu, D. (1977), \textit{On the elimination of nuisance parameters}, JASA \textbf{72}, 355-366.
\bibitem{Berger99} Berger, J. O., Liseo, B., and Wolpert, L. (1999), \textit{Integrated likelihood methods for eliminating nuisance parameters}, Statistical Science \textbf{14}, 1-28.
\bibitem{Bish75} Bishop, Y., Fienberg, S. and Holland, P.(1975), \textit{Discrete Multivariate Analysis, Theory and Practice}, Cambridge, Massachusetts: MIT Press.
\bibitem{Bolfarine92} Bolfarine, H., Leite, J. G. and Rodrigues, J.(1992), \textit{On the Estimation of the Size of a Finite and Closed Population}, Biometrical Journal \textbf{34}, 577-593.
\bibitem{Castle81} Castledine, B. J. (1981), \textit{A Bayesian Analysis of Multiple Recapture }, JASA \textbf{81}, 338-346.
\bibitem{Chadra49} ChandraSekar, C. and Deming, W.E. (1949), \textit{On a method of estimating birth and death rates and the extent of registration},  JASA \textbf{44}, 101-115.
\bibitem{Chao00} Chao, A., Chu, W. and Chiu, H.H.(2000), \textit{Capture-Recapture when Time and Behavioral Response Affect Capture Probabilities}, Biometrics \textbf{56}, 427-433.
\bibitem{Chatterjee14} Chatterjee, K. and Mukherjee, D. (2014), \textit{On the Estimation of Population Size from a Complex Dual-record System}, in arXiv:1408.2153v2 [stat.ME]; \textit{arxiv.org/abs/1408.2153}.
\bibitem{Cox75} Cox, D. R. (1975), \textit{Partial likelihood}, Biometrika \textbf{62}, 269-276.
\bibitem{Cox-Reid87} Cox, D. R. and Reid (1987), \textit{Parameter orthogonality and approximate conditional inference (with discussions)}, J. R. Statist. Soc. B \textbf{49}, 1-39.
\bibitem{George90} George, E. I. and Robert, C. P.(1990), \textit{Capture-recapture models and Bayesian sampling}, Technical Report No. 435, Department of Statistics, Stanford University, Stanford, California.
\bibitem{George92} George, E. I. and Robert, C. P.(1992), \textit{Capture-recapture estimation via Gibbs sampling}, Biometrika, \textbf{79}, 677-683.
\bibitem{Green75} Greenfield, C. C. (1975), \textit{On the estimation of a missing cell in a 2 x 2 contingency table}, J. R. Statist. Soc. A \textbf{138}, 51-61.
\bibitem{Hugg89} Huggins, R.(1989), \textit{On the statistical analysis of capture-recapture experiments}, Biometrika, \textbf{76}, 133-140.
\bibitem{Lee98} Lee, S. M. and Chen, C.W.S.(1998), \textit{Bayesian inference of Population Size for behavioral response models}, Statistica Sinica, \textbf{8}, 1233-1247.
\bibitem{Lee03} Lee, S. M., Hwang, W.H. and Huang, L.H.(2003), \textit{Bayes estimation of Population Size from Capture-recapture Models with Time Variation and Behavior response}, Statistica Sinica, \textbf{13}, 477-494.
\bibitem{Lloyd94} Lloyd, C.J.(1994), \textit{Efficiency of martingle methods in recapture studies}, Biometrika, \textbf{81}, 305-315.
\bibitem{Nour82} Nour, E. S. (1982), \textit{On the Estimation of the Total Number of Vital Events with Data from Dual-record Collection Systems}, J. R. Statist. Soc. A \textbf{145}, 106-116.
\bibitem{Otis78} Otis, D.L., Burnham, K.P., White, G.C. and Anderson, D.R.(1978), \textit{Statistical Inference from Capture Data on Closed Animal Populations}, Wildlife Monographs, \textbf{62}, 1-135.
\bibitem{Raj77} Raj, D. (1977), \textit{On Estimating the Number of Vital Events in Demographic Surveys}, JASA \textbf{72}, 377-381.
\bibitem{Robert67} Roberts, H . V.(1967), \textit{Informative stopping rules and inferences about population size}, JASA, \textbf{62}, 763-775.
\bibitem{Salasar14} Salasar, L. E. B., Leite, J. G. and Louzada, F. (2014), \textit{On the integrated maximum likelihood estimators for a closed population capture–recapture model with unequal capture probabilities}, Statistics, 2014. DOI:10.1080/02331888.2014.960870
\bibitem{Severini98} Severini, T.A. (1998), \textit{An approximation to the modified profile likelihood function}, Biometrika, \textbf{85}, 403-411.
\bibitem{Severini00} Severini, T.A. (2000), \textit{Likelihood Methods in Statistics}, Oxford University Press Inc., New York.
\bibitem{Smith88} Smith, P.J. (1988), \textit{Bayesian Analysis for multiple capture-recapture surveys}, Biometrics, \textbf{44}, 1177-1189.
\bibitem{Smith91} Smith, P.J. (1991), \textit{Bayesian Analysis for a multiple capture-recapture model}, Biometrika, \textbf{78}, 399-407.
\bibitem{Wolter86} Wolter, K. M. (1986), \textit{Some Coverage Error Models for Census Data}, JASA, \textbf{81}, 338-346.
\bibitem{Xu14} Xu, Y., Fyfe, M., Walker, L. and Cowen, L. L. (2014), \textit{Estimating the number of injection drug users in greater Victoria, Canada using capture-recapture methods}, Harm Reduction Journal, \textbf{11:9}, 1-7.
\end{thebibliography}
\end{document}